# The Probability of an Eigenvalue Number Fluctuation in an Interval of a Random Matrix Spectrum

M. M. Fogler and B. I. Shklovskii

*Theoretical Physics Institute, University of Minnesota, 116 Church St. Southeast, Minneapolis, Minnesota 55455*

(October 11, 1994 4:11pm)

We calculate the probability to find exactly $n$ eigenvalues in a spectral interval of a large random $N \times N$ matrix when this interval contains $s \ll N$ eigenvalues on average. The calculations exploit an analogy to the problem of finding a two-dimensional charge distribution on the interface of a semiconductor heterostructure under the influence of a split gate.

Random-matrix theory found its applications in numerous branches of physics. Among them there are the statistical theory of slow neutron resonances, the theory of chaotic systems, the properties of small metallic samples (mesoscopic physics) *etc.* [1]. Random matrices were proven to be a good approximation for the Hamiltonians of such systems.

The random-matrix theory studies statistical properties of spectra of large matrices whose elements have a given random distribution. Much interest attracted the properties of a spectral interval containing on average $s$ eigenvalues, where $s$ is a small fraction of total number of eigenvalues $N$. The actual number $n$ of eigenvalues in this interval fluctuates from one realization of a random matrix to another. Among different statistical characteristics of these fluctuations, two particular ones were the objects of intensive study. One of them was the variance of $n$: $\langle \delta n^2 \rangle = \langle (n-s)^2 \rangle$. Using a thermodynamical argument Dyson [2] showed that

$$\langle \delta n^2 \rangle = \frac{2}{\pi^2 \beta} \ln s + B_\beta, \qquad (1)$$

where $\beta = 1, 2$, or 4 for the three possible ensembles of random Hamiltonians: orthogonal, unitary, and symplectic, respectively. The constant $B_\beta$ was calculated by Dyson and Mehta [3]. It is different for different ensembles, but is generally of order unity. Here we quoted Dyson's result in its simplest form when possible degeneracies and series of non-interacting levels are neglected. The fact that $\langle \delta n^2 \rangle \ll s$ is the consequence of a so-called level repulsion.

Another popular quantity was the probability density $P(d)$ to find two consecutive levels separated by a distance of $d$ average level spacings. For the orthogonal ensemble $P(d)$ is well approximated by the famous "Wigner surmise" $P(d) = (\pi d/2) \exp(-\pi d^2/4)$ [4] if $d$ is not too large. The level repulsion can be seen in the fact that $P(d) \to 0$ as $d \to 0$. For large $d$ the asymptotic behavior of $P(d)$ is different from that of "Wigner surmise" [2,1]:

$$\ln P(d) = -\frac{\pi^2}{16} d^2 - \frac{\pi}{4} d + O(\ln d) \qquad (2)$$

In this paper we calculate analytically the asymptotical behavior of a more general and more informative quantity. This quantity designated by Mehta [1] as $E_\beta(n, s)$ is the probability to find exactly $n$ eigenvalues in a spectral interval containing $s$ eigenvalues on average. It is easy to see that $P(s) = d^2 E_\beta(0, s)/ds^2$ and $\langle \delta n^2 \rangle = \sum_{n=0}^{\infty} (n-s)^2 E_\beta(n, s)$. The idea behind our calculation is to push further the analogy of the problem at hand to the thermodynamics of a Coulomb gas. It was this analogy that Dyson used to obtain Eq. (2). As an important result we would like to present the expression for $E_\beta(n, s)$ in the limit $|n - s| \ll s$:

$$\ln E_\beta(n, s) = -\frac{\pi^2 \beta}{4} \frac{\delta n^2}{\ln(8s/|\delta n|) + B} + O(\ln |\delta n|), \quad (3)$$

where $\delta n = n - s$ and $B$ weakly depends on $\delta n$. This distribution is nearly Gaussian for $|\delta n| \ll \ln s$, and Eq. (1) follows. The Gaussian form agrees with the conjecture put forward by Al'tshuler *et al.* [5]. For $|\delta n| \gg \ln s$ the probability of a fluctuation $\delta n$ in the eigenvalue number is, however, significantly smaller than as it would be determined from a Gaussian distribution with the variance given by Eq. (1).

Let us now turn to the derivation of the general expression for $E_\beta(n, s)$. Suppose we consider the Gaussian orthogonal ensemble $E_{1G}$ defined as the ensemble of real symmetrical matrices $H$ such that their elements $H_{ij}$ are statistically independent and the probability $Q(H)dH$ that a system of $E_{1G}$ will belong to the volume element $dH = \prod_{i \leq j} dH_{ij}$ is invariant under real orthogonal transformations

$$Q(H')dH' = Q(H)dH, \quad H' = S^T H S \qquad (4)$$

for any real orthogonal matrix $S$. The joint probability density function for the eigenvalues $x_i$ of $H$ has the form [1]

$$Q(x_1, \ldots, x_N) = C_N \prod_{i<j} |x_i - x_j| \exp\left[-\sum_{i=1}^{N}(ax_i - b)^2\right], \qquad (5)$$

The quantities $a, b$, and $C_N$ are some constants. Eq. (5) can be rewritten as

$$Q(x_1, \ldots, x_N) = C \exp(-W/T), \quad (6)$$

where

$$W = \sum_i (ax_i - b)^2 - \sum_{i<j} \ln |x_i - x_j| \quad (7)$$

and $T = 1$. Hence, the probability $Q$ is nothing else as corresponding Gibbs' factor for a gas of $N$ point charges in thermodynamic equilibrium at temperature $T$. Each pair of charges $i$ and $j$ exhibits electrostatic repulsion logarithmically dependent on the distance $|x_i - x_j|$ between them. The $N$ charges are confined by an external potential $\phi_c(x) = (ax - b)^2$. Once this analogy is established one can apply the methods of the classical thermodynamics to calculate various statistical quantities of random matrix spectra.

The generalization of Eq. (6) to the other two types of ensembles requires only change in the value of the temperature: $T = 1/\beta$. Introduce the partition function

$$\psi_\beta = \frac{1}{N!} \int \ldots \int e^{-\beta W} dx_1 \ldots dx_N. \quad (8)$$

The factor $1/N!$ corresponds to our treatment of charges as indistinguishable particles [2]. The quantity $E_\beta(n, s)$ is the ratio of two values of $\psi_\beta$ calculated for two different domains of the integration. The first domain is determined by the condition that exactly $n$ charges belong to the given interval of length $s$. The second value corresponds to the unrestricted domain of integration.

When both $n$ and $s$ are large one attempts to abandon the discreet formulation of the problem in favor of the continuum one. Following Dyson we make three assumptions: (i) there is a macroscopic eigenvalue density function $\rho(x)$; (ii) for a given $\rho(x)$ the free energy of the gas is composed of two parts

$$F = V_1 + V_2, \quad (9)$$

where $V_1$ is the potential energy in the mean-field approximation

$$V_1 = -\frac{1}{2} \int \int \rho(x)\rho(y) \ln |x - y| dx dy + \int \rho(x)\phi_c(x) dx \quad (10)$$

and $V_2$ is the contribution depending only on the local density

$$V_2 = \int \rho(x) f_\beta[\rho(x)] dx, \quad (11)$$

$f[\rho]$ being the free energy per particle of a Coulomb gas with average density $\rho$ in equilibrium; (iii) the overwhelmingly dominant contribution to the integral (8) comes from configurations not deviating significantly from an "optimal" density fluctuation, namely that function $\rho(x)$ which makes $F$ a minimum subject to

$$\int_{-t}^{t} \rho(x) dx = n, \quad (12)$$

where $t = s/2$.

Another simplification may be obtained by noting that if $s \ll N$ then we study only a small fraction of the whole spectral interval. In this case the explicit form of the confinement potential in the expression (7) is irrelevant and can be chosen according to our needs. We find it the most convenient to think that the confinement is due to a compensating background of unit density. The appropriate modification of Eq. (10) is then

$$V_1 = -\frac{1}{2} \int \int [\rho(x) - 1][\rho(y) - 1] \ln |x - y| dx dy \quad (13)$$

As Dyson showed [2] the free energy density $f[\rho]$ is

$$f[\rho] = (1/\beta - 1/2)\rho \ln \rho. \quad (14)$$

The first term is just the entropy multiplied by the temperature and the second term is the correlation energy per unit length.

The probability $E_\beta(n, s)$ is expressed as

$$\ln E_\beta(n, s) \approx -\beta \min F, \quad (15)$$

where $F$ is given by Eqs. (9, 11, 13, 14) and the minimum is sought in the class of continuous non-negative functions $\rho(x)$ under the condition (12). This optimal fluctuation is very close to the one, which ensures the lowest value of the mean-field part $V_1$ of the free energy $F$. However, the optimization have to be done taking into account the local part $V_2$ as well, and we will make an error of order $O(\ln |\delta n|)$ as long as $s \gg 1$ and $|\delta n| \gg 1$. As discussed by Dyson [2] we lost such order terms anyway upon the transition from the discreet to the continuum formulation of the problem.

Using the variational principle we find that the electrostatic potential

$$\phi(x) = -\int \rho(y) \ln |x - y| dy \quad (16)$$

created by the optimal density fluctuation must satisfy the following conditions

$$\begin{aligned} \phi(x) &= -V, & |x| < t, \; \rho(x) > 0 \\ &> -V, & |x| < t, \; \rho(x) = 0 \\ &= 0, & |x| > t \end{aligned} \quad (17)$$

if $n < s$ and similarly,

$$\phi(x) = V, \quad |x| < t, \quad \rho(x) > 0$$
$$< V, \quad |x| > t, \quad \rho(x) = 0 \qquad (18)$$
$$= 0, \quad |x| > t$$

otherwise. The meaning of these conditions is that the electron gas breaks into "metallic" regions, where the potential is perfectly screened, and "insulating" regions, where the charge density $\rho(x)$ vanishes and there is no screening.

Consider the case $n < s$ first. We notice that this problem is equivalent to another one, which at first glance appears completely different. We are referring to a problem of finding the charge distribution of a laterally confined two-dimensional electron gas (2DEG) in a heterostructure. The reason for this is as follows. Due to a large aspect ratio in some of such devices, they may be considered as translationally invariant in one of the dimensions. Then the usual Coulomb interaction of two-dimensional charges leads to the logarithmic interaction in terms of the one-dimensional charge density. This is precisely the type of interaction that we are having in our Coulomb gas model.

Consider a simplified model of the split-gate device on the GaAs/Al$_x$Ga$_{1-x}$As heterostructure studied in Ref. [6]. In this model (see Fig. 1), ionized donors and the two-dimensional electron gas are characterized by continuous charged densities. The donors constitute a uniform positive background compensated by the 2DEG. The split gate is represented by two semi-infinite metal planes separated by the gap of width $2t$ centered at $x = 0$. The system is translationally invariant in $\hat{y}$-direction and all the charges and the gate are in the plane $z = 0$.

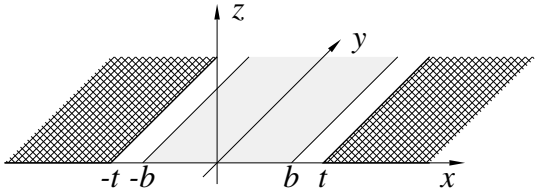

FIG. 1. The model of the split-gate device. The two shaded areas represent the gate in the shape of two semi-infinite planes. Beneath the gate the two-dimensional electron gas located in $z = 0$ plane is depleted. The density of the electron gas does not vanish only in a narrow strip (grey area) between the two halves of the gate.

Suppose a voltage difference $V$ is applied between the gate and the 2DEG. The 2DEG will be depleted under the gate and near its edges. Therefore, near the edges of the gate the charge density is due to the background only. If we subtract the total charge density from the background charge density, then this new quantity will be non-zero in three disconnected regions: in a central strip of 2DEG and in the metallic gate. Having in mind the analogy discussed above, we designate this quantity by the *same symbol* $\rho(x)$ as the eigenvalue density of a random matrix. Let the width of the strip of the 2DEG be $2b$, then the $\rho(x)$ in units of the background charge density is [6]

$$\rho(x) = \mathrm{Re}\sqrt{(b^2 - x^2)/(t^2 - x^2)}. \qquad (19)$$

The consequence for the distribution of the eigenvalue density in the optimal fluctuation is now that it is clearly given by the same formula (19). The parameter $b < t$ is to be found from Eq. (12). The graph of $\rho(x)$ is shown in Fig. 2(a). Note that for $n = s$ $b = 0$ and $\rho(x) = 0$ in the interval $-t < x < t$. This particular case was studied by Dyson [2] to obtain the asymptotic form of $E_\beta(0, s)$ and $P(s)$.

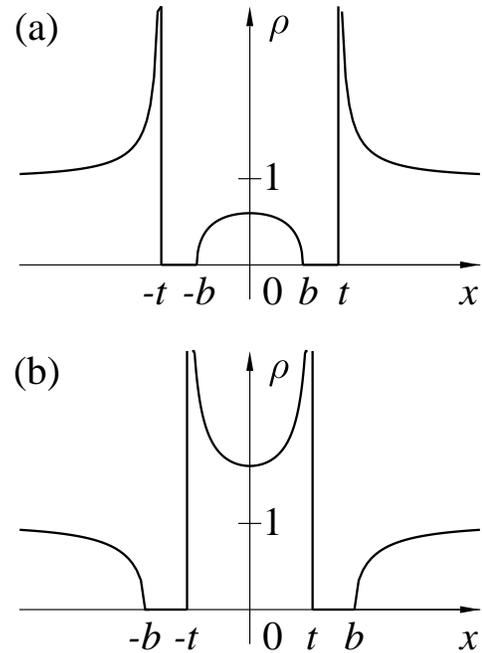

FIG. 2. The optimal eigenvalue density fluctuation for (a) $n < s$ (b) $n > s$.

To calculate the free energy corresponding to the fluctuation we need to know the distribution of the electric field in the Coulomb gas model. It can also be obtained knowing the corresponding distribution for the split-gate device. Namely, the field in the Coulomb gas is by the factor of two weaker. The reason for this is that the potential due to real two-dimensional charges is $-2\int \rho(y) \ln|x - y|$, whereas the potential (16) in the Coulomb gas model is twice as small. With this correction the electric field in the Coulomb gas model is given by the expression

$$-d\phi(x)/dx = \pi\,\mathrm{Im}\sqrt{(b^2 - x^2)/(t^2 - x^2)}, \qquad (20)$$

The electric field vanishes everywhere where $\rho(x)$ is nonzero in agreement with Eq. (17).

The case $n > s$ can be reduced to a similar electrostatic problem with the gate in the shape of an infinite strip of width $2t$ to which a positive voltage $V$ is applied. The eigenvalue density and the electric field in the Coulomb gas are given by the same expressions (19,20) but now $b > t$ and the regions of zero $\rho(x)$ are formed outside the interval $|x| < t$ [see Fig. 2(b)].

The calculation of the parameters of the optimal fluctuations is now straightforward and the answer can be given in terms of the complete elliptic integrals [7]:

$$V_1 = \frac{\pi^2 t^2}{4} - \frac{Vn}{2}$$
$$V = \pi t [E(k) - k'^2 K(k)]$$
$$n = 2t[E(k') - k^2 K(k')] \qquad (21)$$

when $n < s$ and

$$V_1 = -\frac{\pi^2 t^2}{4}\frac{k^2}{k'^2} + \frac{Vn}{2}$$
$$V = \frac{\pi t}{k'}[K(k) - E(k)]$$
$$n = \frac{2t}{k'}E(k') \qquad (22)$$

when $n > s$. Here, $k^2 = 1 - k'^2$ and $k'$ is defined as $k' = b/t$ for $n < s$ and $k' = t/b$ for $n > s$. Also $V_2 = V$ in both cases. Eq. (15) now reads

$$\ln E_\beta(n,s) = -\beta V_1 - (1 - \beta/2)V + O(\ln|n-s|). \qquad (23)$$

To use this formula one have to first find the $k$ corresponding to given $n$ and $s$ and then substitute this value into the expressions for $V$ and $V_1$. For example, for $n = 0$ one finds that $k = 1$ and then $V = \pi t$ and $V_1 = \pi^2 t^2/4$. One observes that our formulae indeed reproduce the result Eq. (2) upon the substitution $d = 2t$.

Consider now the case of small fluctuations: $|\delta n| \ll s$, which corresponds to $k \ll 1$. In this limit the expansion of Eqs. (21,22) in asymptotic series yields

$$V_1 = \frac{\pi^2}{8}(tk^2)^2\left[\ln\frac{4}{k} + \frac{1}{4} + O(k^2)\right]$$
$$V = \frac{\pi}{2}tk^2[1 + O(k^2)]$$
$$|\delta n| = tk^2\left[\ln\frac{4}{k} + \frac{1}{2} + O(k^2)\right]. \qquad (24)$$

Using Eq. (23) one recovers Eq. (3). As a possible method of the numerical check of the latter, we suggest calculating the ratio of two subsequent even moments $M_{2m+2}$ and $M_m$ of $E_\beta(n,s)$ defined as $M_m = \sum_{n=0}^\infty (n-s)^m E_\beta(n,s)$. This ratio is expected to be

$$\frac{M_{2m+2}}{(2m+1)M_2 M_m} \simeq 1 - \frac{1}{2}\frac{\ln 2am}{\ln s}, \qquad (25)$$

where $a = 2/\pi^2\beta$ and $1 \ll am \ll s^2$. In contrast to the same kind of ratio for the Gaussian distribution, where it is just equal to unity, in our case it decreases with $m$. For instance, at $m \sim s/2a$ it should be about 0.5.

For completeness we provide also the asymptotic form of $\ln E_\beta(n,s)$ for *large* positive $\delta n$, i.e., for $n \gg s$:

$$\ln E_\beta(n,s) = -\beta n^2 \left(\ln\frac{8n}{\pi s} - \frac{3}{2}\right)$$
$$- (1 - \beta/2)n\left(\ln\frac{8n}{\pi s} - 1\right) + O(\ln n). \qquad (26)$$

One observes that the main terms in both Eq. (3) and Eq. (26) are quadratic in $\delta n$ and contain large logarithmic factors. This can be explained in terms of the electrostatic analogy as follows. The quantity $\ln E_\beta(n,s)$ is up to a constant the work required to charge a two-dimensional capacitor by $\delta n$ charge units. This work is equal to $\delta n^2/2C$, where $C$ is the capacitance. In the case of $|\delta n| \ll s$ the width $s$ of the central plate of this capacitor is much larger than the gap between the plates and $C$ is logarithmically large. In the case of $n \gg s$ the situation with geometrical parameters of the capacitor is reversed, and its capacitance is thus inversely proportional to a large logarithmic factor.

In conclusion, we studied the probability $E_\beta(n,s)$ to find a given number $n$ of eigenvalues in an interval of a random matrix spectrum defined by the condition that this interval contains $s$ eigenvalues on average. We calculated the asymptotical behavior of $E_\beta(n,s)$ for large $s$. It is found to be Gaussian for small fluctuations of $n$ around its average value and to decay faster than Gaussian at $|\delta n| \gtrsim \ln s$. We suggested a method of the numerical verification of our result based on calculating of large order moments of $E_\beta(n,s)$.

This work was supported by NSF under Grant No. DMR-9321417.